\documentclass{article}
\usepackage[english]{babel}
\usepackage[margin=1in]{geometry}
\usepackage{background}
\usepackage[math]{blindtext}% just to generate filler text

\SetBgScale{4}
\SetBgColor{gray}
\SetBgAngle{90}
\SetBgContents{}
\SetBgPosition{-1.5,-10}

\begin{document}

% -- NJP: 
%\title[Observation of the sling effect]{Observation of the sling effect} 

{\centering

{\bfseries\Large Observation of the sling effect\bigskip}

% -- NJP
%\author{Gregory P. Bewley$^{1}$}

%\author{Ewe-Wei Saw$^{1}$}

%\author{Eberhard Bodenschatz$^{1,2}$}
%\address{$^1$Max Planck Institute for Dynamics and Self-Organization, 
%37077 G\"{o}ttingen, Germany}
%\address{$^2$Laboratory of Atomic and Solid State Physics 
%and Sibley School of Mechanical and Aerospace Engineering, 
%Cornell University, Ithaca, NY 14853, USA}

%\ead{\mailto{gregory.bewley@ds.mpg.de}}
% -- NJP

Gregory P. Bewley\textsuperscript{1} , Ewe Wei Saw\textsuperscript{1} , Eberhard Bodenschatz\textsuperscript{1,2} \\
  {\itshape
\textsuperscript{1}Max Planck Institute for Dynamics and Self-Organization, 
37077 G\"{o}ttingen, Germany \\
\textsuperscript{2}Laboratory of Atomic and Solid State Physics 
and Sibley School of Mechanical and Aerospace Engineering, 
Cornell University, Ithaca, NY 14853, USA \\
\normalfont (Published August 27, 2013 in NJP)

  }
}

\begin{abstract}

When cloud particles are small enough, they move with the turbulent air in the cloud.  
On the other hand, as particles become larger their inertia affects their motions, 
and they move differently than the air.  
These inertial dynamics impact cloud evolution and ultimately climate prediction, 
since clouds govern the earth's energy balances.  
Yet we lack a simple description of the dynamics.  
Falkovich et al.\,describes theoretically 
a new dynamical mechanism called the ``sling effect'' 
by which extreme events in the turbulent air cause 
idealized inertial cloud particles to break free from the airflow 
(Falkovich G, Fouxon A, Stepanov MG 2002 \textit{Nature} \textbf{419} 151).  
The sling effect thereafter causes particle trajectories to cross each other 
within isolated pockets in the flow, 
which increases the chance of collisions 
that forms larger particles.  
We combined experimental techniques 
that allow for precise control of a turbulent flow 
with three-dimensional tracking of multiple particles at unprecedented resolution.  
In this way, we could 
observe both the sling effect and crossing trajectories between real particles.  
We isolated the inertial sling dynamics 
from those caused by turbulent advection 
by conditionally averaging the data.  
We found the dynamics to be universal in terms of a local Stokes number 
that quantifies the 
local particle velocity gradients.  
We measured the probability density of this quantity, 
which shows that sharp gradients become more frequent 
as the global Stokes number increases.  
We observed that sharp compressive gradients in the airflow initiated the sling effect, 
and that thereafter gradients in the particle flow ran away and 
steepened in a way that produced singularities in the flow in finite time.  
During this process both the fluid motions and gravity became unimportant.  
The results underpin a framework for describing a crucial aspect of inertial particle dynamics 
and predicting collisions between particles.  
\bigskip

%\noindent PACS numbers: 06.20.Jr, 95.30.Dr, 95.30.Sf, 98.62.Ra, 98.80.-k, 98.80.Es, 98.80.Jk
\noindent{\it Keywords}: sling effect, inertial particles, rain initiation, clouds, turbulence

\end{abstract}

% -- NJP: 
%\pacs{}
%\vspace{2pc}
%\noindent{\it Keywords}: sling effect, inertial particles, rain initiation, clouds, turbulence

%\submitto{\NJP}
% -- NJP

\section{Introduction}
% INTRODUCTION

Clouds are water particles dispersed by a complex turbulent air flow, 
and are the primary source of uncertainty in the prediction of the climate~\cite{heintzenberg:2009}.  
Underlying the uncertainty is lack of understanding of cloud particle size evolution, 
which is mediated by a strong coupling between air turbulence 
and microphysical processes~\cite{bodenschatz:2010}.  
In particular, turbulence controls the growth and decay of the particles that compose the cloud.  
The growth of the particles is caused in part by the collision 
and coalescence of smaller particles~\cite{devenish:2012}.  

The turbulent air in clouds carries water particles back and forth 
on scales as small as a millimeter and as large as the cloud itself.  
When cloud particles move in unison with the surrounding air, 
collisions occur simply because the particles are big enough 
to run into each other 
as the air slides past itself~\cite{saffman:1956}, 
though it is well known that this happens infrequently.  
In this description, both 
the air flow and the particle flow are incompressible.  
In fact the particle flow is compressible~\cite{maxey:1987}, 
even when the air flow is not.  
The compressibility is due to the 
particles' inertia, which results in their concentrating locally, 
thereby increasing the probability of their collisions~\cite{eaton:1994,balkovsky:2001}.  
This, however, is still not sufficient to fully capture the physics of the particle motions.  
According to Falkovich et al.\,\cite{falkovich:2002}, 
there is a qualitatively distinct effect that might explain the rate at which particles collide.  
They suggest that the particle flow is not only compressible 
but also interpenetrating, 
so that groups of particles go through each other.  
This prediction agrees with our experiments.

\begin{figure}[h]
\centering
\includegraphics[width=7.75cm]{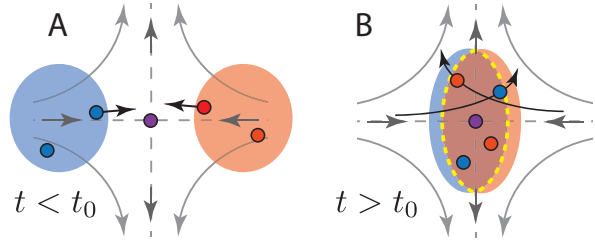}
\caption{
Illustration of the sling effect.  
In a small region in a cloud, an intense turbulent motion of the air 
(grey arrows) 
compresses groups of particles together 
(red and blue ovals).  
(A) Initially, the particle velocities vary smoothly from the left to the right.  
These conditions persist if 
the air damps the particle motions to the extent that their trajectories never cross.  
The particles then turn away from each other as does the air.  
If, however, their approach is fast enough, 
particles coming from different directions go past each other.  
This is a sling.  
(B) The particles' inertia carries them so far 
that they create a pocket, 
the region between the dashed yellow lines, 
where particles are more likely to hit each other because they cross paths, 
rather than moving alongside each other.  
Within the pocket, the particles' velocities fall in three classes (at least), 
according to the branch of the multi-valued field they belong to.  
Loosely, the particles move either 
to the right (blue), left (red), or stand still (purple).  
Eventually, the particles all relax toward the air's motion, 
which itself evolves.  
}
\label{fig:schematic}
\end{figure}

The interpenetration of particles is thought to occur in isolated volumes that we call pockets, 
which form in locally compressive airflows, or stagnation points.  
This makes the sling effect intrinsically a small-scale phenomenon.  
It arises on scales where the airflow is smooth 
and its variations approximately linear.  
Within pockets, 
nearby particles move in different directions, 
as seen in Fig.\,1.  %\ref{fig:schematic}.  
Outside of the pockets, nearby particles 
trace approximately parallel trajectories~\cite{lanterman:2004}.  
While the latter behavior is a matter of common experience, 
the interpenetration of trajectories has until now not been seen, 
except in a two-dimensional flow~\cite{denissenko:2006}.  

Our first result is the experimental observation of crossing trajectories
between particles with nearly the same size.  
Figure~2 %\ref{fig:examples} 
shows such trajectories 
reconstructed from images of particle shadows 
simultaneously acquired by two cameras~\cite{maas:1993,malik:1993,ouellette:2006}, 
as described in the methods section.  
The dimension of the visible region is 
only a few Kolmogorov scales, $\eta$, of the turbulence, 
which required the development of a particle tracking system with high resolution.  
Nearby trajectories are observed to cross, 
as they would within a pocket.  
When they cross, the particles on different trajectories are about 2$\eta$ apart.  
These trajectories are entirely incompatible 
with those of fluid particles at this scale.  
Note that the crossing of particle trajectories with each other is distinct from 
the crossing of particle trajectories over fluid trajectories~\cite{yudine:1959}, 
which can be caused by inertia or by gravitational settling.  
These phenomena have been studied numerically~\cite{hill:2005} 
and experimentally~\cite{wells:1983}.  
As we shall see, the sling effect causes the trajectories of particles with the same size to cross, 
even in the absence of gravity.  

We confirm the existence of the pockets through the singular dynamics 
that necessarily precede their formation.  
These dynamics were named the sling effect by Falkovich et al.\,\cite{falkovich:2002}.  
The development of a pocket is illustrated in Fig.\,1.  %\ref{fig:schematic}.  
The singularity that arises, which defines the boundary of the pocket, 
is called a caustic~\cite{wilkinson:2005}, 
and is analogous to 
features in the patterns that light forms on the bottom of a pool of water~\cite{berry:1981}.  
Such pockets cannot form in the air itself because the air cannot go through itself, 
but they can form in the particle field of a cloud.  
Collisions are more likely within pockets, because the particles cross paths, 
as in a busy intersection, 
rather than moving along side each other, as in a highway on-ramp.

\begin{figure}
\centering
\includegraphics[width=15cm]{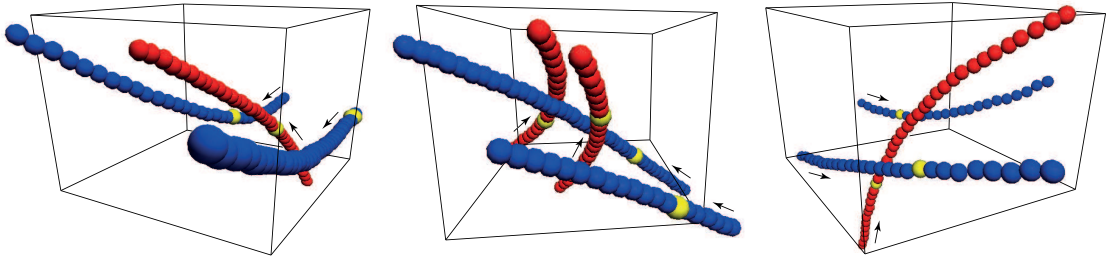}
\caption{
Slinging particles.  
The spheres mark the positions of 19\,$\mu$m particles at intervals of 67\,$\mu$s ($\approx \tau_\eta$/33) 
in a frame with dimensions of about 0.7\,$\times$\,0.7\,$\times$\,0.5\,mm 
($\approx$ 4$\eta$\,$\times$\,4$\eta$\,$\times$\,3$\eta$).  
For low Stokes numbers, particles followed each other through the volume 
along approximately parallel trajectories.  
Here, the Stokes number was 0.50, 
and we see particles moving in two distinct directions, 
distinguished by the two colours.  
The motions correspond to those of particles in a pocket, as seen in Fig.\,1B.  %\ref{fig:schematic}B.  
%Because the motions are damped through friction with the air, the trajectories aligned over time.  
The spheres are about twice the actual size of the particles relative to the frames.  
The yellow spheres mark the positions of the particles at a given time.  
}
\label{fig:examples}
\end{figure}

The singular dynamics of slings can be seen qualitatively in 
simplified equations of motion for 
particles whose mass density, $\rho_p$, is much higher than that of the surrounding air, 
and whose diameters, $d$, are much smaller than all scales of motion of the air~\cite{tchen:1947}: 
$d \mathbf{v} / d t = (\mathbf{u} - \mathbf{v}) / \tau_p  +  \mathbf{g}$, 
where the particles are tracers of the velocity field $\mathbf{v}(\mathbf{x}, t)$, 
the air moves with velocities $\mathbf{u}(\mathbf{x}, t)$, 
the particle response time is $\tau_p = \rho_p d^2 / 18 \mu$, 
the viscosity of the air is $\mu$, 
the acceleration of gravity is $\mathbf{g}$, 
and $d/dt = \partial/\partial t + \mathbf{v} \cdot \nabla$.  
The gradient of this equation is 
$d \sigma / dt = (s - \sigma)/\tau_p - \sigma^2, $
where $\sigma = \nabla \mathbf{v}$ and $s = \nabla \mathbf{u}$ 
are $3\times3$ matrices, 
and $\sigma^2$ is the square of the 
matrix $\sigma$.  
Because $\mathbf{g}$ drops out of the equation, 
we see that the sling effect is not gravitational.  
We will now see that operating with the gradient 
isolates a specific role of inertia, 
embodied in the nonlinear term, $\sigma^2$.  

For one longitudinal component, 
$\sigma_{11} = \partial v_1 / \partial x_1$, 
which is a scalar and not a matrix, 
the particle velocity gradient equation reads 
\begin{equation}
\frac{d \sigma_{11}}{dt}
 = (s_{11} - \sigma_{11}) / \tau_p
 - \sigma_{11}^2
 - \sigma_{12}\sigma_{21}
 - \sigma_{13}\sigma_{31}.  
\label{eq:slingeq}
\end{equation}
Informally, 
when $|s_{11}|$ and $|\sigma_{11}|$ are small relative to $1/\tau_p$, 
the $\sigma^2$ terms can be neglected, 
and the evolution of $\sigma_{11}$ is controlled by $s_{11}$.  
That is, the particles move approximately with the air flow.  
The key point is that if $|\sigma_{11}|$ 
is large relative to $1/\tau_p$, 
then the $\sigma_{11}^2$ term can drive the system toward a singularity in finite time.  
To see this, 
consider a coordinate system that maximizes $\sigma_{11}^2$.  
Because this term is nonnegative and much larger than 
$\sigma_{11}/\tau_p$, 
it can lead to divergence to minus infinity 
if $\sigma_{11}$ is initially negative.  
The singularity thereafter evolves into a pocket, 
within which the field description of the particle dynamics breaks down 
because $\mathbf{v}$ becomes multivalued~\cite{falkovich:2002,wilkinson:2005}.  
In other words, in a strong enough compressive gradient, 
the particle motions transition from being over-damped to under-damped.  
In contrast, if $\sigma_{11}$ is initially large and positive, 
corresponding to a violent expansion, 
the nonlinearity enhances the damping of the expansion; 
the particles move away from each other, the flow smooths out, and no sling occurs.  
A sling is thus a self-generating extreme event in the particle field, 
which was suggested by Falkovich et al.\,\cite{falkovich:2002} 
to be initiated by the very intermittent turbulent fluctuations in a cloud.  

In order for the sling framework to be relevant to cloud particle size evolution, 
the turbulent fluctuations have to have the right properties to initiate slings.  
In other words, 
we need to determine experimentally whether the conditions are right in the flow 
for slings to occur.  
The fluctuations in $s_{11}$ need to be both strong enough and persistent enough 
to make $-\sigma_{11}$ large relative to $1/\tau_p$, 
and so to push the particles together sufficiently violently.  
Furthermore, 
they must not thereafter neutralize the slings by blowing the particles back apart.  
The resulting pockets must be so sparse that the 
field description of the particle motions, 
embodied in Eq.\,\ref{eq:slingeq}, 
is still useful for describing their formation.  
On the other hand, 
the pockets must occur sufficiently often 
for the slings to be the main explanation of collisions, 
and the number of collisions occurring in each pocket must be sufficiently high 
to generate a larger number of collisions than 
those produced by other mechanisms.  

Evidence in favor 
of the sling effect, and 
of there being frequent sling-induced collisions, 
comes from 
numerical models~\cite{falkovich:2002,wilkinson:2005,falkovich:2007,ducasse:2009,meneguz:2011}.  
These studies focused on the number of collisions produced by slings, 
or on the spatial structure that slings create.  
One key aspect of our experimental study is the focus on the 
the dynamical nature of the theory, which has not received attention.  
Most importantly, the numerical models were simplified to the extent of the theory itself, 
a point we now pursue.  

Equation~\ref{eq:slingeq} is necessarily an imperfect representation of reality.  
To illustrate this, consider that it treats a continuum of non-interacting 
and identical point particles 
that experience linear Stokes drag and do not affect the motions of the background airflow, 
whereas real particles do not satisfy these conditions.  
Real particles are different from each other 
and are more naturally thought of as moving individually.  
The description of the particles does not include any of the effects, such as the history, 
added mass, or nonlinear drag forces, that real particles experience~\cite{maxey:1983}.  
The history term, for example, 
may suppress caustics~\cite{daitche:2011}.  
The motions of real particles also distort the airflow, 
and the particles interact with each other hydrodynamically~\cite{balanchandar:2010}.  
Finally, the equation is only valid as long as the particle velocity field is single-valued, 
which breaks down exactly when the subject of interest, the sling effect, occurs.  

Due to limitations in computational power, 
numerical simulations were limited until now to 
the same simplifying assumptions as the theory, 
embodied in Eq.~\ref{eq:slingeq}.  
When additional physics were taken into account, 
the resolution of the background flow suffered, 
as was the case in Daitche and T\'el~\cite{daitche:2011} 
where the flow was chaotic rather than turbulent.  
The various additional effects 
experienced by real particles, 
and the possible interactions between them, 
are so numerous 
that they call for an experiment to identify the sling effect itself.  

In our experiment, we observed dense populations of liquid droplets 
with nearly the same size in a turbulent airflow.  
As described quantitatively in the methods section, 
the liquid was much more dense than the air 
and the droplets were much smaller than the smallest scales of turbulent motion.  
The particles were weakly inertial, 
in the sense that their Stokes numbers were smaller than one.  
The Stokes number, $St$, quantifies the relevance of the dropletsÕ inertia, 
and is $St = \tau_p/\tau_\eta$, 
where $\tau_\eta$ 
is the characteristic time scale of the smallest and fastest turbulent motions.  
As discussed in the supplementary information, 
gravity had only a small influence on the particle dynamics.  

% THE EVIDENCE

In order to show that the sling effect produced pockets like those in Fig.~2, %\ref{fig:examples}, 
we used two- and three-particle statistics to characterize the particle velocity field.  
With two particles observed at the same time, we estimated 
$\sigma_{11} \approx \delta \mathbf{v}(\mathbf{r}, t) \cdot \mathbf{r}(t) / r^2$ 
in a coordinate system aligned with the separation vector, $\mathbf{r}$.  
Because the turbulence was isotropic, 
we combined the $\sigma_{11}$ statistics 
measured when $\mathbf{r}$ pointed in different directions.  
Here 
$\delta \mathbf{v}(\mathbf{r}, t) = \mathbf{v}(\mathbf{x}+\mathbf{r}, t) - \mathbf{v}(\mathbf{x}, t)$ 
was the velocity difference between the two particles.  
This quantity is a good approximation of the gradient 
when $r$ is smaller than the length scales of the field $\mathbf{v}$, 
but not where there are singularities in $\sigma$.  
In our analysis, 
we restricted $r$ to be smaller than $3\eta$.  
As discussed below and in the supplementary information, 
the average separation was about $2\eta \approx 0.4$\,mm, 
which was more than 10 times larger than the particles' diameters.  
Such small separations are difficult to resolve for two reasons.  
First, particles can be found at a given separation increasingly 
infrequently as the separation gets smaller.  
Second, it is challenging for both the optics and the image-analyzing 
software to discriminate between and to track particles that are only a few diameters apart.  

Our experiments demonstrate that slings indeed occur.  
The particles qualitatively moved as if they were in pockets, 
the particle velocity field had large-enough gradients for slings to occur, 
the large negative gradients tended to grow increasingly sharp over time, and 
the turbulent fluctuations in the air did not 
prevent evolution toward a singularity in finite time.  

\begin{figure}
\centering
\includegraphics[width=7.25cm]{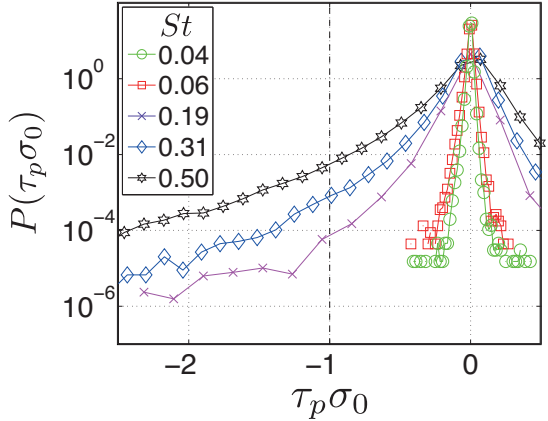}
\caption{
The probability density of particle field gradients, $\tau_p \sigma_0$.  
Roughly speaking, 
a gradient must be smaller than $-1$
in order for a sling to happen.  
This happened 
with measurable frequency 
when the Stokes number, $St$, was large enough.  
In both this figure and in Fig.~\ref{fig:grads}, 
the gradients were computed for particles separated by up to 3$\eta$.  
The legend shows the values of the Stokes number for the various data sets.  
}
\label{fig:gradprob}
\end{figure}

\section{Results}

To provide insight into the results in Fig.~2, 
we study the dynamics of velocity gradients.  
Figure~3 %\ref{fig:gradprob} 
shows the 
probability density 
of the particle velocity gradients.  
The tails of the distributions grow markedly broader with increasing Stokes number, 
and the distributions are skewed toward negative gradients.  
We focus here on the negative tails, 
corresponding to compressive gradients.  

The compressive gradients 
exceeded $1/\tau_p$ with measurable probability 
when the Stokes number was 0.19, 0.31 and 0.50, but not when the Stokes number was 0.04 or 0.06.  
This indicates that the sling theory applied for high enough Stokes numbers.  
At the highest Stokes number, 0.50, 
these large gradients accounted for about 0.17\% 
of the measurements.  
This percentile is the integral of the probability density from minus infinity to minus one.  
For the Stokes number 0.30 and 0.19 cases, 
the fractions were 0.018\% and 0.0016\%, respectively, 
which shows that the domain where the sling theory applied 
shrank quickly as the Stokes number decreased.  
In each case, these probabilities were 
low enough that the field theory 
probably 
applied for most of the data, and in particular, 
for those regions just preceding the formation of pockets, as we show below.  
Note that these are the probability densities of the measurements, 
which are not the same as the probability densities of the underlying velocity fields 
since the particles did not sample the fields uniformly.  

We identified the formation of singularities 
by measuring the rate of change of the gradients, 
$d \sigma_{11}/dt$.  
Negative rates of change drive $\sigma_{11}$ to minus infinity through the sling effect.  
We did not observe individual particles for long enough 
to trace the evolution toward singularities in individual realizations.  
Therefore, we consider the statistical quantities 
$\sigma_0$ and $s_0$, 
which are the averages of $\sigma_{11}$ and $s_{11}$ conditioned on small intervals of $\sigma_{11}$.  
These averages also satisfy Eq.\,\ref{eq:slingeq}, since the intervals are small.  
We show in the supplementary information that 
the $\sigma_{12}\sigma_{21}$ and $\sigma_{13}\sigma_{31}$ terms were negligible 
when $|\tau_p \sigma_0|$ was large, 
which is the focus of our analysis.  
During a sling, the $\sigma_0^2$ term dominates, 
so that a graph of $d \sigma_0/dt$ versus $\sigma_0$ 
is an upside-down parabola.  
The parabola is shifted to the left by the damping term, $-\sigma_0$, 
and to the right by the driving term, $s_0$.  

\begin{figure}
\centering
\includegraphics[width=15cm]{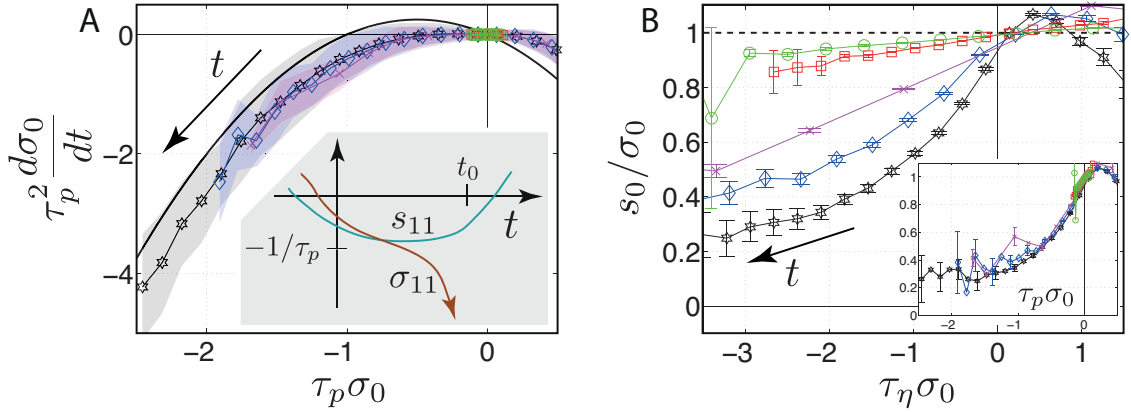}
\caption{
Sling dynamics.  
(A)
Negative, compressive gradients in the particle field 
grew increasingly sharp 
in a way that led in finite time to singularities and pockets.  
The illustration in the inset shows an evolution of gradients 
that would reproduce the curves traced by the data.  
This divergent behavior was visible only when the Stokes number was large enough, 
and even then was uncommon, 
with the frequency of measurements made for each $\tau_p \sigma_0$ 
being given in Fig.~\ref{fig:gradprob}.  
The shaded regions 
cover data within one standard deviation of the mean.  
The solid curve, 
$\tau_p^2 d\sigma_0 / dt = -(\tau_p \sigma_0 + \tau_p^2 \sigma_0^2)$, 
traces particles slinging through still air.  
It captures the essential features of the data, 
with the deviations being attributable to motions of the air.  
(B)
Where the particles moved with the air, 
the ratio shown here 
of conditionally averaged air and particle velocity gradients 
was nearly one.  
This held only for small gradients and small Stokes numbers.  
During a sling, the particle gradients diverge, 
so that the ratio between particle and air gradients approaches zero, 
which is indeed what we observed.  
The inset shows that the local Stokes number of the particle flow, $\tau_p \sigma_0$, 
determines the decoupling of the particles from the airflow.  
%  data collapse under normalization by $\tau_p$, 
%the particle response time.  
The error bars 
mark the 95\% confidence intervals for the means.  
The legend is the same as for Fig.\,3.  
}
\label{fig:grads}
\end{figure}

Figure~4A %\ref{fig:grads}A 
shows that the data indeed fall on a parabola 
so that large negative gradients 
grew increasingly negative.  
These are the dynamics of slings, 
that is, of nascent singularities or caustics.  
The parabolic shape of the data, with $d\sigma_0/dt \sim -\sigma_0^2$, implies that 
the time until the divergence of the gradients was given by 
their initial values, up to a logarithmic correction.  
That is, the gradients diverged approximately as $(t-t_0)^{-1}$, 
where $t_0$ is the time at which the singularity formed.  
After $t_0$, pockets like the ones in Figs.\,1 and 2 %\ref{fig:examples} and \ref{fig:schematic} 
formed~\cite{falkovich:2002,wilkinson:2005}.  
There is a systematic deviation in Fig.~4A 
between the data and the theoretical model, 
which we interpret in what follows.  

As a sling progresses, particle velocity gradients grow arbitrarily large relative to the gradients in the air.  
On the other hand, if turbulent fluctuations conspired to prevent the creation of pockets, 
the gradients in the air would counteract a newly initiated sling by changing sign to push the particles back apart.  
To see which of the two possibilities happened in the experiment, 
we measured the ratio $s_0/\sigma_0$, 
which could either approach zero or change sign.  
We did not measure $s_0$ directly, 
but evaluated its ratio with $\sigma_0$ 
through Eq.\,\ref{eq:slingeq}, so that it was approximately equal to 
$(d \sigma_0 / dt + \sigma_0^2) (\tau_p / \sigma_0)  + 1$.  
As shown in the Materials and Methods section, 
the $\sigma_{12}\sigma_{21}$ and $\sigma_{13}\sigma_{31}$ terms 
cancel and do not need to be neglected in this calculation.  

Fig.\,4B %\ref{fig:grads}B 
exposes that the particles moved with the air when their 
inertia was negligible, since the gradients in the air and in the particle field were nearly equal.  
As can be seen from Fig.\,3, 
most particles moved in this regime.  
When the particles' inertia was important, 
the ratio of gradients, $s_0/\sigma_0$ approached zero, 
meaning that the particle gradients steepened much more quickly 
than the gradients in the background airflow.  
This shows that the particle dynamics decoupled from the airflow dynamics, 
and it distinguishes the dynamics as inertial particle dynamics and not airflow dynamics.  
The ratio did not change sign, 
which indicates that turbulent fluctuations did not prevent the formation of singularities.  
Note that the difference between the ratio and zero quantifies the deviation of the data 
from the model in Fig.~4A.  

We found that one parameter, $\tau_p \sigma_0$, 	
captured 
the transition from the particles moving with the air 
to decoupling from it.  
As seen in the inset to Fig.\,4B, the data with different Stokes numbers fall on a single curve 
when plotted against $\tau_p \sigma_0$.  
This parameter is itself a Stokes number 
based on the local particle flow gradients, $\sigma_0$, 
rather than on the characteristic airflow gradient, $1/\tau_\eta$, and so 
can be called a local Stokes number.  
The key point is that knowledge of the particle flow 
was sufficient to explain the particle dynamics.  
The parameter 
also tells us how to predict the typical particle field gradient, 
given knowledge of the airflow, and independent of the global Stokes number.  
In other words, it can help predict where in space slings will occur.  

What is intrinsic in our analysis of the sling effect is that the essential flow geometry 
is the compressive flow, 
or stagnation point, 
as illustrated in Fig.~1.  
Stagnation points produce normal stresses, such as $\sigma_{11}$, 
and it is these normal stresses that dominate Eq.\,\ref{eq:slingeq} during a sling 
and not the off-diagonal terms, such as $\sigma_{12}$, which capture rotation in the fluid.  
The normal stresses, in particular, 
predicted in our data the decoupling of the particle flow from the airflow, 
as seen in Fig.\,4B.

% DISCUSSION AND CONCLUSION
\section{Conclusions}

Our results demonstrate that the sling effect happens in clouds.  
It occurs alongside other physics such as gravitational settling.  
While differential settling and differential inertia 
are effective when particles with different sizes encounter one another~\cite{shaw:2001}, 
the sling effect provides a way for particles to collide \emph{even when they are the same size}.  
Quantifying the relative importance of the various phenomena is a matter of ongoing study.  

The importance of the sling effect 
is that it provides a way to organize thinking about collisions.  
The pockets are distinct objects 
and have properties independent of the turbulence 
that predict the number of collisions they produce~\cite{gustavsson:2011}.  
As a matter of principle, the pockets  
rule out an accurate description of the particle flow by a single-valued field.  
The global Stokes numbers for cloud droplets 
are lower than those for which 
we observed the sling effect experimentally, 
but it is the \emph{local} Stokes number that is the relevant parameter.  
We expect slings and pockets to become more prevalent 
as the Reynolds number increases, 
as extreme events in the flow become more common.  %\cite{}.  
In particular, 
their prevalence in clouds should be higher than in our experiment 
because Reynolds numbers in clouds are much larger than those in our experiment.  

Our results apply also to particles in other settings 
and provide, for example, bounds on when they can be used 
as tracers of rare events in turbulence~\cite{toschi:2009}.  
Taken together, this information is crucial to the correct modeling of particle collisions specifically 
and to the science of extreme events in fluids broadly.  

%% == end of paper:

\section{Methods}
% METHODS

We performed the experiments with a so-called ``soccer-ball'' apparatus, 
which is described in detail in the supplementary information.  
Here we provide an overview of our methods.  
The flow-generating apparatus produced statistically stationary and isotropic turbulence.  
It was similar to the one described by Chang et al.\,\cite{chang:2012}.  
The apparatus consisted of an acrylic shell with an inner diameter of 1\,m, 
which contained air at standard temperature and pressure.  
Holes in the shell accommodated thirty-two loudspeakers, 
the particle generator, and the imaging systems.  
The loudspeakers pushed and pulled air through conical nozzles to form air jets, 
which pointed toward the centre of the chamber 
where they interacted with each other~\cite{hwang:2004}.  
The jets had the same strength, 
so that the turbulence was approximately isotropic, 
the residual anisotropy being less than 5\%, 
as measured by $u'_{y}/u'_{z}$.  
The turbulence had neither a significant mean shear nor a mean flow.  

We observed liquid water-alcohol particles in the turbulent air flow.  
Particles of two sizes, 6.8
%$\pm$2 
and 19\,$\mu m$, 
%$\pm$4
with standard deviations of 2 and 4\,$\mu m$, respectively, 
were injected by a spinning disk~\cite{walton:1949} 
until their number density was about 1000~$cm^{-3}$.  
The advantage of the spinning disk is that it produced large numbers of droplets 
while maintaining a narrow size distribution relative to sprays.  
The number density of the larger particles 
corresponded to a volume fraction of $4\times10^{-6}$
and a mass fraction of $3\times10^{-3}$ relative to the surrounding air.  
These values are on the border where particles start to influence turbulence.  
The presence of higher densities of particles is known to reduce the intensity of turbulence, 
but broadly speaking, the effect of particles on turbulence is not known~\cite{balanchandar:2010}.  
We draw from this that while the particles in our experiment 
may have had a small influence on the airflow, 
the effect was not strong enough to prevent us from observing the sling effect.  

The turbulent energy dissipation rate per unit mass, $\epsilon$, 
was varied to produce different particle Stokes numbers.  
The Stokes number, as defined above, 
is $St = \tau_p / \tau_\eta$, 
where $\tau_\eta = \sqrt{\nu / \epsilon}$ 
is the Kolmogorov time scale of the turbulence, 
$\nu = \mu/\rho_f$ is the viscosity of the air, 
and $\rho_f$ is the mass density of the air.  
The Stokes number 0.06 and 0.50 cases corresponded to flows 
where $\epsilon =$~3.2$\pm$0.2\,$m^2/s^3$, 
for the Stokes number 0.04 and 0.31 cases, 
$\epsilon =$~1.2$\pm$0.5\,$m^2/s^3$, 
and for the Stokes number 0.19 case, 
$\epsilon =$~0.45$\pm$0.5\,$m^2/s^3$.  
We determined $\epsilon$ from the dissipation-range scaling of the 
longitudinal second-order structure functions.  
That is, we fit to measurements of the second-order structure function 
a power law to determine the unknown coefficient, $\epsilon$.  
The Stokes numbers were low enough not to affect substantially 
the structure functions in the dissipation range~\cite{bec:2010,salazar:2012}.  
These dissipation rates corresponded respectively to Kolmogorov length scales, 
$\eta = (\nu^3/\epsilon)^{1/4}$, 
of 180, 230 and 300\,microns, 
which are the scales over which the velocity gradients were approximately linear, 
and to Taylor Reynolds numbers 190, 170 and 160, 
which signify that the flow was fully turbulent~\cite{frisch:1995}.  

We used white light sources to produce shadows of the particles 
on the sensors of two digital movie cameras.  
That is, the particles in a volume at the center of the ball were back-lit, 
so that the light passed through the volume directly into the cameras.  
The main advantage of shadow imaging is that much less light is required 
than when images of scattered light are desired.  
A second advantage is that we could measure the sizes of the droplets 
from the images of the droplet shadows.  
Movies of the particle shadows were recorded synchronously by the two cameras.  
Each camera pixel imaged $3.3\,\mu m\times3.3\,\mu m$ of real space, 
which corresponded to a pixel resolution of about $\eta/50$.  
The frame rate was 15\,kHz, which was faster than $30/\tau_\eta$.  
The cameras captured 512$\times$512 pixels, 
so that the view volume was about $2\,mm^3$.  
The volume was also controlled by limiting the depth-of-field of the camera lenses 
to about $2\,mm$.  
Particles that were outside of this depth-of-field were simply blurred out.  

To find the 3D positions of the particles in real space, 
we combined the two-dimensional coordinates of the particles' images 
in two cameras~\cite{ouellette:2006b}.  
From each realization of the experiment, 
the data consisted of particle positions and sizes at a series of times.  
We sorted these data by particle size, and split the data into two parts.  
The large particles and small particles were thereafter processed separately.  
Particles were tracked in three-dimensions using a three-frame predictive algorithm, 
whereby particle accelerations were minimized~\cite{ouellette:2006}.  

To measure the rate of change of the gradients, 
we estimated the gradients at times $t_1$ and $t_2 = t_1 + \Delta t$ with the same pair of particles.  
For each pair of particles that we observed, we fixed the coordinate system 
to the particle separation vector $\mathbf{r}(t)$ at time $t_1$, 
since the velocity gradient equation, Eq.\,\ref{eq:slingeq}, applies in an inertial frame of reference.  
A single pair of particles then yields the gradient estimate 
\begin{equation}
Q(t)
= \frac{ \delta \mathbf{v}(\mathbf{r}, t) \cdot \mathbf{\hat{r}}(t_1) }
{ \mathbf{r}(t) \cdot \mathbf{\hat{r}}(t_1) }, 
\end{equation}
where $\mathbf{\hat{r}}(t)$ was a unit vector in the direction of $\mathbf{r}(t)$.  
For times $|t-t_1|$ small relative to $r/|\delta \mathbf{v}|$, 
and for separations $r$ small relative to the length scales of the velocity field, 
the quantity is an accurate estimate of the gradient, so that $Q(t_1) \approx \sigma_{11}(t_1)$.  
For our data, $\Delta t$ was always more than 8 times smaller than $r/|\delta \mathbf{v}|$, 
and $r$ was always smaller than 3$\eta$.  
Because the flow was isotropic, 
we combined the measurements for different values of $t_1$ 
and for all possible particle pairs.  
The rate of change of $Q$ is related to the rate of change of the gradient 
by the equation 
\begin{equation}
\frac{d \sigma_{11}}{d t} 
= \frac{d Q}{d t} 
- \sigma_{12} \sigma_{21} 
- \sigma_{13} \sigma_{31}, 
\label{eq:sigandQ}
\end{equation}
where the cross terms arise because of the rotation of the particle pair.  
The derivation of this equation is similar to the one of 
Li and Meneveau~\cite{li:2005}, the main difference being that in 
their analysis, the coordinate system rotates with the particle pair, 
while in ours it does not.  
We evaluated the rate of change of $Q$ by a finite difference, 
so that 
\begin{equation}
\frac{d Q}{d t} \approx \frac{Q(t_2) - Q(t_1)}{\Delta t}.  
\end{equation}
In our analysis of the dynamics of the gradients, $\sigma_{11}$, 
we neglected the cross terms, $\sigma_{12} \sigma_{21}$ and $\sigma_{13} \sigma_{31}$, 
so that $d \sigma_{11} / d t \approx d Q / d t$.  
This was justified because the cross terms were small.  
We describe how we checked this
in the supplementary information.  
Note that the cross terms arise both in Eq.\,\ref{eq:sigandQ}, 
and on the right side of Eq.\,\ref{eq:slingeq}, 
so that they cancel in the evolution equation for $Q$.  
It follows that our determination of $s_0$, shown in Fig.\,4B, 
is valid to the extent that Eq.\,\ref{eq:slingeq} is, 
and is not affected by neglecting the cross terms.  
That is, $s_0/\sigma_0 = (d Q / dt + \sigma_0^2) (\tau_p / \sigma_0)  + 1$.

\section{Acknowledgements}
We thank Tobias Schneider and J\"{u}rgen Vollmer for discussion, 
and Poh Yee Lim for help operating the experiment.

%\section*{References}


\begin{thebibliography}{10}

\bibitem{heintzenberg:2009}
Heintzenberg J and Charlson RJ
\newblock 2009 \textit{Clouds in the perturbed climate system: 
  their relationship to energy balance, atmospheric dynamics, and precipitation}
\newblock (MIT Press)

\bibitem{bodenschatz:2010}
Bodenschatz E, Malinowski SP, Shaw RA and Stratmann F
\newblock 2010 Can we understand clouds without turbulence? 
\newblock \textit{Science} \textbf{327} 970--971

\bibitem{devenish:2012}
Devenish BJ, {et~al.}
\newblock 2012 Droplet growth in warm turbulent clouds
\newblock \textit{Q. J. R. Meteorol. Soc.} \textbf{138} 1401--1429

\bibitem{saffman:1956}
Saffman PG and Turner JS
\newblock 1956 On the collision of drops in turbulent clouds
\newblock \textit{J. Fluid Mech.} \textbf{1} 16--30

\bibitem{maxey:1987}
Maxey MR
\newblock 1987 Gravitational settling of aerosol particles in homogeneous
  turbulence and random flow fields. 
\newblock \textit{J. Fluid Mech.} \textbf{174} 441--465

\bibitem{eaton:1994}
Eaton JK and Fessler JR
\newblock 1994 Preferential concentration of particles by turbulence
\newblock \textit{Int. J. Multiphase Flow} \textbf{20} 169--209

\bibitem{balkovsky:2001}
Balkovsky E, Falkovich G and Fouxon A
\newblock 2001 Intermittent distribution of inertial particles in turbulent flows
\newblock \textit{Phys. Rev. Lett.} \textbf{86} 2790--2793

\bibitem{falkovich:2002}
Falkovich G, Fouxon A and Stepanov MG
\newblock 2002 Acceleration of rain initiation by cloud turbulence
\newblock \textit{Nature} \textbf{419} 151--154

\bibitem{lanterman:2004}
Lanterman DD, {et~al.}
\newblock 2004 Characterizing intense rotation and dissipation in turbulent flows
\newblock \textit{Chaos} \textbf{14} S8

\bibitem{denissenko:2006}
Denissenko P, Falkovich G and Lukaschuk S
\newblock 2006 How waves affect the distribution of particles that float on a liquid surface
\newblock \textit{Phys. Rev. Lett.} \textbf{97} 244501

\bibitem{maas:1993}
Maas HG, Gruen A and Papantoniou D
\newblock 1993 Particle tracking velocimetry in three-dimensional flows, part 1. 
  photogrammetric determination of particle coordinates
\newblock \textit{Exp. Fluids} \textbf{15} 133--146

\bibitem{malik:1993}
Malik NA, Dracos T and Papantoniou DA
\newblock 1993 Particle tracking velocimetry in three dimensional flows, part 2. 
  particle tracking
\newblock \textit{Exp. Fluids} \textbf{15} 279--294

\bibitem{ouellette:2006}
Ouellette NT, Xu H and Bodenschatz E
\newblock 2006 A quantitative study of three-dimensional {L}agrangian
  particle tracking algorithms
\newblock \textit{Exp.\ Fluids} \textbf{40} 301--313

\bibitem{yudine:1959}
Yudine MI
\newblock 1959 Physical considerations on heavy-particle diffusion
\newblock \textit{Adv. Geophys.} \textbf{6} 185--191

\bibitem{hill:2005}
Hill RJ
\newblock 2005 Geometric collision rates and trajectories of cloud droplets falling 
  into a Burgers vortex
\newblock \textit{Phys. Fluids} \textbf{17} 037103

\bibitem{wells:1983}
Wells MR and Stock DE
\newblock 1983 The effects of crossing trajectories on the dispersion of particles 
  in a turbulent flow
\newblock \textit{J. Fluid Mech.} \textbf{136} 31--62

\bibitem{wilkinson:2005}
Wilkinson M, Mehlig B
\newblock 2005 Caustics in turbulent aerosols
\newblock \textit{Europhys. Lett.} \textbf{71} 186--192

\bibitem{berry:1981}
Berry MV
\newblock 1981 Singularities in waves and rays
\newblock \textit{Les Houches Lecture Series Session XXXV} \textbf{35} 453--543

\bibitem{tchen:1947}
Tchen CM
\newblock 1947 Ph.D. thesis (Technische Hooge School Delft).

\bibitem{falkovich:2007}
Falkovich G and Pumir A
\newblock 2007 Sling effect in collisions of water droplets in turbulent clouds
\newblock \textit{J. Atmos. Sci.} \textbf{64} 4497--4505

\bibitem{ducasse:2009}
Ducasse L and Pumir A
\newblock 2009 Inertial particle collisions in turbulent synthetic flows:
  Quantifying the sling effect
\newblock \textit{Phys. Rev. E} \textbf{80} 066312

\bibitem{meneguz:2011}
Meneguz E and Reeks MW
\newblock 2011 Statistical properties of particle segregation in homogeneous
  isotropic turbulence
\newblock \textit{J. Fluid Mech.} \textbf{686} 338--351

\bibitem{maxey:1983}
Maxey MR and Riley JJ
\newblock 1983 Equation of motion for a small rigid sphere in a nonuniform flow
\newblock \textit{Phys. Fluids} \textbf{26} 883--889

\bibitem{daitche:2011}
Daitche A and T\'{e}l T
\newblock 2011 Memory effects are relevant for chaotic advection of inertial particles
\newblock \textit{Phys. Rev. Lett.} \textbf{107} 244501

\bibitem{balanchandar:2010}
Balachandar S and Eaton JK
\newblock 2010 Turbulent dispersed multiphase flow
\newblock \textit{Annu. Rev. Fluid Mech.} \textbf{42} 111--33

\bibitem{shaw:2001}
Shaw RA and Oncley SP
\newblock 2001 Acceleration intermittency and enhanced collision kernels in turbulent clouds
\newblock \textit{Atmos. Res.} \textbf{59--60} 77--87

\bibitem{gustavsson:2011}
Gustavsson K and Mehlig B
\newblock 2011 Distribution of relative velocities in turbulent aerosols
\newblock \textit{Phys. Rev. E} \textbf{84} 045304

\bibitem{toschi:2009}
Toschi F and Bodenschatz E
\newblock 2009 Lagrangian properties of particles in turbulence
\newblock \textit{Annu. Rev. Fluid Mech.} \textbf{41} 375--404

\bibitem{chang:2012}
Chang K, Bewley GP and Bodenschatz E
\newblock 2012 Experimental study of the influence of anisotropy on the
  inertial scales of turbulence
\newblock \textit{J. Fluid Mech.} \textbf{692} 464--481

\bibitem{hwang:2004}
Hwang W and Eaton JK
\newblock 2004 Creating homogeneous and isotropic turbulence without a mean flow
\newblock \textit{Exp. Fluids} \textbf{36} 444--454

\bibitem{walton:1949}
Walton WH and Prewett WC
\newblock 1949 The production of sprays and mists of uniform drop size by
  means of spinning disc type sprayers
\newblock \textit{Proc. Phys. Soc. B} \textbf{62} 341--350

\bibitem{bec:2010}
Bec J, Biferale L, Cencini M, Lanotte AS and Toschi F
\newblock 2010 Intermittency in the velocity distribution of heavy particles in turbulence
\newblock \textit{J. Fluid Mech.} \textbf{646} 527--536

\bibitem{salazar:2012}
Salazar JPLC and Collins LR
\newblock 2012 Inertial particle relative velocity statistics in homogeneous
  isotropic turbulence
\newblock \textit{J. Fluid Mech.} \textbf{696} 45--66

\bibitem{frisch:1995}
Frisch U
\newblock 1995 \textit{Turbulence: The Legacy of A.N. Kolmogorov}
\newblock (Cambridge University Press)

\bibitem{ouellette:2006b}
Ouellette NT, Xu H, Bourgoin M and Bodenschatz E
\newblock 2006 An experimental study of turbulent relative dispersion models
\newblock \textit{New J. Phys.} \textbf{8} 109

\bibitem{li:2005}
Li Y and Meneveau C
\newblock 2005 Origin of non-{G}aussian statistics in hydrodynamic turbulence
\newblock \textit{Phys. Rev. Lett.} \textbf{95} 164502

\end{thebibliography}
\end{document}